# Thermal Alteration and Differential Sublimation Can Create Phaethon's "Rock Comet" Activity and Blue Color

C.M. Lisse[1] and J.K. Steckloff [2,3]

[1] Space Exploration Sector, Johns Hopkins University Applied Physics Laboratory, 11100 Johns Hopkins Rd, Laurel, MD USA 20723  carey.lisse@jhuapl.edu

[2]Planetary Science Institute, Tucson, AZ 85719  jordan@psi.edu

[3]Department of Aerospace Engineering and Engineering Mechanics, University of Texas at Austin, Austin, TX, 78712

14 Pages, 4 Figures, 1 Table

**Abstract**

In 2010 Jewitt and Li published a paper examining the behavior of "comet-asteroid transition object" 3200 Phaethon, arguing it was asteroid-like in its behavior throughout most of its orbit, but that near its perihelion, at a distance of only 0.165 AU from the sun, its dayside temperatures would be hot enough to vaporize rock (>1000 K, Hanus *et al.* 2016). Thus it would act like a "rock comet" as gases produced from evaporating rock were released from the body, in a manner similar to the more familiar sublimation of water ice into vacuum seen for comets coming within ~3 AU of the Sun. In this Note we predict that the same thermal effects that would create "rock comet" behavior with $Q_{gas} \sim 10^{22}$ mol/sec at perihelion have also helped to greatly bluen Phaethon's surface via preferential thermal alteration and sublimative removal of Fe and refractory organics, known reddening and darkening agents. These predictions are testable by searching for signs of spectral bluening of the surfaces of other small bodies in Phaethon-like small perihelion orbits, including comets, and by *in situ* measurements of Phaethon's surface and coma composition near perihelion with the upcoming DESTINY+ mission (Kawakatsu & Itawa 2013, Arai *et al*. 2018) to Phaethon by JAXA.

**I. Introduction.** 3200 Phaethon (1983 TB), is an active Apollo asteroid (Jewitt & Hsieh 2006, Hanus *et al.* 2016, Jewitt & Li 2010, Jewitt *et al.* 2013, 2018) and parent body of the Geminids meteor shower of mid-December. It is on an 1.434 yr period, highly eccentric (e = 0.88990) orbit that brings it within 0.14 AU of the Sun, closer than any other named asteroid (MacLennan *et al.* 2021 and references therein). Identified in the 1983 IRAS sky survey (Green & Davies 1983, Green *et al.* 1985) as the first asteroid ever discovered from space and classified as a rare B-subtype of the C-type asteroids (Tholen 1984, Kartashova *et al.* 2021), it has now been well characterized over 25+ orbits and is known to be an object ~6 km in diameter, rotating at about 3.6 hrs' period, with a highly unusual blue color vs. other asteroids (Jewitt & Hsieh 2006, Hanus *et al.* 2016, Kareta *et al.* 2018).

Phaethon's mass shedding is an important observable that needs to be put in context. A thermal driving mechanism for its mass loss into the Geminid stream has been proposed and studied by numerous recent authors (Jewitt & Li 2010, Jewitt *et al.* 2013, 2019; Li & Jewitt 2013; Tabeshian *et al.* 2019, Nagano *et al.* 2020, Takir *et al.* 2020, MacLennan *et al.* 2021, Masiero *et al.* 2021, Ye *et al.* 2021). Blaauw 2017 states an estimated minimum total mass for the Geminid meteoroid stream of $1.6 \times 10^{16}$ g, while JPL Horizons 2019 quotes an estimated mass for Phaethon itself of $1.4 \times 10^{17}$ g, only 8.75 times larger. So Phaethon has lost at least 12% of its mass into fragments following its orbit. It is not clear if Phaethon is continually shedding material to resupply the Geminids or created the fragments during an impulsive phase a few Myr ago, soon after its most recent transition into its current close q = 0.143 AU orbit (MacLennan *et al.* 2021). By contrast, the mass required to optically resurface Phaethon out to 5 um would be only about 10 um (depth) * 3 g cm$^{-3}$ (density) * $4\pi R_{Phaethon}^2$ (surface area) ~ $5 \times 10^9$ g, about one millionth of the mass shed into the Geminid stream.

Phaethon's surface composition is by all accounts of carbonaceous chondrite material that has been heavily thermally altered. Its surface spectrum is ***extremely*** and unusually blue (Licandro et al. 2007, DeMeo et al. 2009, Clark *et al.* 2013), increasing ~30% in reflectivity from 1.0 um down to 0.5 um (Fig. 1); in fact it is one of the bluest asteroids known. Licandro *et al.* (2007) found the best meteorite spectral matches for Phaethon's spectrum to be Yamato-86720, an unusual CI/CM meteorite that likely has been heated up to 500–600 ºC (Hiroi *et al.* 1996), and heat-treated Ivuna (Hiroi *et al.* 2004), while Clark *et al.* (2013) found the best spectral matches to be the thermally processed CK4 meteorites ALH85002 and EET 92002. The CK4 meteorites are known for their highly oxidized, CAI, olivine, refractory metal sulfide, FeO$_x$ plus highly refractory inorganic amorphous carbon/graphite matrix dominated composition due to extensive hot aqueous reworking of primitive ferromagnesian silicates, sulfides, and complex organics.

Thus it is very plausible that the heating Phaethon endures during its perihelion passage has led to compositional alteration of its sensible surface, an alteration involving less than a millionth of the mass it has lost into space within the last few Myrs. In this paper we make the case that any asteroidal body on a Phaethon-like orbit with perihelion < 0.15 AU is likely to have a bluened surface, due to preferential destructive sublimation and removal of any complex organic (i.e., CHON-containing) and nanophase Fe phases faster than the ~Myr timescales over which these materials are created by solar wind ion implantation and space weathering surface alteration processes (Nesvorny *et al.* 2005, Lazzarin *et al.* 2006). While it is possible that Phaethon is intrinsically blue like the handful of other known main belt B-type asteroids (Clark *et al.* 2013), and thermal processing is causing ongoing mass wasting every perihelion passage and thus merely continually exposing fresh new blue surface, we find this circumstance highly fortuitous and unlikely. Our prediction that ***ANY*** C-type asteroid on a Phaethon-like low-q orbit will be extremely blue can be tested by observations of a large ensemble of sun-skirting asteroids to search for bluening regardless of asteroid type, and to search for surrounding Fe- and CHON-rich gas comae near perihelion. These predictions can be extended to sungrazing comet nuclei, and we find that archival measurements of 96P/Machholz 1 (q=0.12 au) report it to be anomalously blue in color and depleted in carbon-species emission (Eisner *et al.* 2019). Our predictions will also be testable by surface spectral maps and coma measurements performed by the upcoming DESTINY+ *in situ* mission to Phaethon by JAXA (Kasuga *et al.* 2006; Kawakatsu & Itawa 2013, Arai *et al.* 2018).

**II. Thermophysical Analysis & Sublimation Behavior.** Recently, we produced two papers explaining the current state and early evolution of volatile ices in Kuiper Belt Object Arrokoth (Lisse *et al.* 2021, Steckloff *et al.* 2021a), and another analyzing the recent appearance of large amounts of Fe gas around the T Tauri- like protostar RW Aur A (Lisse *et al.* 2022). Both cases appear to be controlled by thermal sublimation processes. E.g., the ices found on small body Arrokoth today appear dominated by the most refractory, strongly hydrogen bonded species such as water and methanol. Our best model for what is occurring in the RW Aur A system includes the catastrophic disruption via impact of an inwardly migrating Ceres sized differentiated planetesimal with an iron core + rocky mantle, followed by evaporation of the fine fragments of the former core in the 1250-1650 K environment of the systems inner accretion disk where the collision occurred. By contrast, though, we calculated that the fragments of the rocky outer mantle would be stable at the temperatures and Pa-level pressures of the accretion disk.

In order to apply our thermal stability versus evaporation analysis to Phaethon, we need to model the thermophysical processing of Phaethon's surface throughout its orbit, especially around perihelion, when its surface becomes hot enough to boil Fe and Mg-pyroxene (Fig. 2)[1]. This requires that we accurately model surface temperatures across Phaethon's surface. We compute the temperatures across Phaethon's surface using the Many Materials Orbital Sublimation model (MaMOS; Springmann *et al.* 2019; Steckloff *et al.* 2020, 2021b), which employs the method of Steckloff *et al.* (2015) to use conservation of energy, Clausius-Clapeyron, and Knudsen-Langmuir (Langmuir, 1913) relations to compute the surface temperature that balances solar energy with sublimative and radiative heat loss over a spherical surface[2]. The resulting MaMOS Phaethon surface temperatures (Fig. 3) are consistent with recent thermophysical models of the surface (Hanus *et al.* 2016, Tabeshian *et al.* 2019; MacLennan *et al.* 2021, Maserio *et al.* 2021) that have the maximum dayside temperature reaching upwards of 1100 K for a Phaethon with Bond Albedo = 0.043, Emissivity ~ 0.9, and beaming parameter = 0.9, despite its relatively quick rotation rate of 3.6 hrs and high thermal inertia of ~600 +/- 200 $J\ m^{-2}\ K^{-1}\ s^{-1/2}$. The MaMOS temperature also appear consistent with thermal modeling of Phaethon constrained by observations (R. Vervack, private communication, and Vervack *et al.* 2021; Fig. 3).

What do these thermophysical models mean? Using the model curves of Lisse *et al.* 2021 and Steckloff *et al.* 2021, surface temperature profiles translate into local saturation vapor pressures for exposed surface solid species that are highly non-linear in the local surface temperature (Fig. 2). Using MAMOS to compute the surface temperatures and resulting molecular flux across the surface of an idealized spherical Phaethon with dayside to nightside temperatures varying by hundreds of degrees, we integrate across the surface to compute the maximum total sublimative molecular production from Phaethon at each point along its orbit (Fig. 4) and find that Phaethon can lose up to ~$10^{22}$ molecules/s of iron (depending on the $P_{sat}(T)$ literature model used) at perihelion.

I.e., for a $T_{surface\ max}$ = 1080 K model at the sub-solar point (i.e., local noon) at perihelion, we find that the 5.2 x $10^{11}$ $cm^2$ of Phaethon's surface loses about 9 monolayers of Fe on average each day; it takes only about 2.7 hours to remove an entire monolayer, and about 50 monolayers are removed in each orbit. One micron's worth of Fe is removed in only 1000 years, and almost 1 mm in ~1 Myr, much faster than the estimated ~1 Myr timescale to redden a surface via space weathering (Domingue *et al.* 2014; Kohout *et al.*

---

[1] The absolute scaling of the $P_{sat}$ vs T curves for iron and rock can vary by 1-2 orders of magnitude depending on which literature $P_{sat}$ vs. T dependence is adopted, but the relative instability of Fe and pyroxene versus stable olivine leading to removal of remains consistent. The literature needs a thorough modern review paper to clean up and order discrepant published values, as Fray & Schmitt 2009 did for volatile icy materials found in outer solar system bodies.

[2] A simple spherical surface assumption is justified as Hanus *et al.* 2016 showed that because of its relatively high orbital inclination of ~22 deg and obliquity of 148 +/- 10 deg, coupled with solar tidal forces, the hottest sub-solar point of Phaethon wanders over 100 deg of body latitude over ~$10^4$ yrs (~7000 orbits), explaining the lack of any hemispherical photometric color differences found in the observations of Ohtsuka *et al.* (2009) and Kartashova *et al.* (2019).

2014, 2016; Pieters & Noble 2016; Quadery *et al.* 2015; Schelling *et al.* 2015). Thus the optical characteristics of the surface can be changed via thermally driven sublimation.[3]

The same arguments just given are also applicable to Mg-rich pyroxene endmember enstatite, which has similar sublimative temperature dependence to Fe at 500 – 1100K (Fig. 2), yielding a sublimative mass loss rate of pyroxene at the $\sim 10^{22}$ mol/sec level (depending on the $P_{sat}(T)$ literature model used) at perihelion (Fig. 4). It is important to note that pyroxenes thermally decompose as they sublimate, disproportionating into SiO + $O_2$ vapor and leaving behind a solid refractory olivine residue (Tachibana *et al.* 2002). The sublimative loss of bulk enstatite materials can thus weaken the surface solid matrix of Phaethon but leave behind solids that can be shed as entrained dust in the sublimative gas outflow, producing the Geminid meteor stream material detected at Earth.

As for solid materials such as water ice and refractory complex organics, we note that while both are expected to be incorporated in abundance in C-type asteroids, we find that they are much too volatile (Fig. 2) to remain in the surface layers of Phaethon to be of importance.

**III.    Discussion.** We can thus reasonably expect over thousands of years that any solid fine Fe particles, along with any refractory organic solids and the most labile of rocks (like pyroxene) in the surface layers of Phaethon near the sub-solar region (which wanders over the body's surface, Hanus *et al.* 2016) will evaporate. This result provides a natural explanation for the blueness of Phaethon's surface, since the primary finding in studies of space weathered asteroid surfaces is that on timescales of ~1 Myr, solar XUV and solar wind sputtering combine to energetically burn and reduce the surface portions of asteroids while extracting Fe from Fe-olivine and putting it into dark, reddish nanophase Fe particles (Reddy *et al.* 2012; Domingue *et al.* 2014; Kohout *et al.* 2014, 2016, 2020; Quadery *et al.* 2015; Schelling *et al.* 2015; Pieters & Noble 2016; Fazio *et al.* 2018; Chrbolková *et al.* 2021, Young *et al.* 2021). Without dark, reddish nanoscale Fe and organic refractory carbon materials, we expect that surface reflectance spectroscopy of Phaethon will show features predominantly from the most stable and refractory CAI, iron oxide, olivine and inorganic amorphous carbon solid species.

---

[3] Since the submission of this paper, a separate study on the thermal behavior of Phaethon near perihelion leading to emission of volatile Na (sodium) gas and formation of a surrounding Na coma has been published by Masiero *et al.* (2021). The model and ideas presented in Masiero *et al.* (2021) are entirely consistent with those presented in this work, in that if one can evaporate Na from rocky material for an object with local surface maximum temperature at perihelion around 1075 K, then metallic Fe should be labile as well.

One could argue that the changes creating the bluening seen on Phaethon are due to high temperature chemical processing of a carbonaceous chondritic surface. While this may be possible initially, it would not protect the surface from further space weathering; to paraphrase Clark *et al.* 2013, 3200 Phaethon looks like high-temperature processed, heavily oxidized CK meteoritic material that has been kept clean of any nanophase Fe, while relatively spectrally similar ²Pallas looks like its bluening has been dampened by added nanophase Fe. Licandro *et al.* 2007 found similarly that Phaethon's best meteoritic matches are highly thermally processed CI and CM meteorites. Either the initial high-T pulse incurred by Phaethon when it moved into its present low-q orbit (MacLennan *et al.* 2021), coupled with abundant available water, has has chemically destroyed any organic refractories[4] and oxidized any FeNi metal to FeOx and any ferromagnesian silicates to Fe-olivines, or sublimation drove off the bulk organic refractories and FeNi metal and pyroxenes, or both. Subsequent nanophase Fe and CHON compounds created by reductive solar wind space weathering of Fe-olivines and inorganic amorphous carbon/graphite over Myrs were then easily removed and the bluest surface (Fig. 1) maintained via sublimation alone, as water was long gone from the object due to sublimation (Fig 2).

Since we predict total gas production rates of ~$10^{22}$ molecules/s Fe and $10^{21}$ molecules/s of SiO and O for a few days around each perihelion passage (Fig. 4), observers can use our estimates to predict when any sensible Fe/Si/O comae might be detected. This will motivate planning for future Phaethon studies by ground based observatories and *in situ* spacecraft like the upcoming DESTINY+ mission to Phaethon (Kasuga *et al.* 2006; Kawakatsu & Itawa 2013, Arai *et al.* 2018). Since bluening may occur via sublimation alteration of space weathered surfaces as well as by mass wasting of surface solids off-body into the Geminid meteor stream, exposing fresh surface to be thermally altered, future measurement of the amount of near-perihelion Fe coma atoms versus the amount of dust surface area lost will help distinguish the relative importance of the two effects, suggesting a natural series of studies by to test our predictions. Measurement of the dust mass loss rate of Phaethon near perihelion (Li & Jewitt 2013) will also help settle the issue of whether the Geminid meteor stream was created impulsively upon Phaethon's insertion into its present day orbit or is supported by renewed dust emission each perihelion passage, and the efficiency of pyroxene disproportionation in producing weakened surface solids for emission.

**Table 1 – Small airless bodies on highly eccentric periodic orbits with perihelion distances < 0.165 AU**

| Object | Semimajor Axis (AU) | Eccentricity | Inclination (degrees) | Perihelion Distance (AU) | Diameter (km) |
|---|---|---|---|---|---|
| **Phaethon** | 1.27 | 0.889 | 22.3 | 0.140 | ~5.8 |
| **2005 UD** | 1.28 | 0.872 | 28.7 | 0.163 | ~1.28 |
| **2021 PH27** | 0.462 | 0.712 | 31.9 | 0.133 | ? |
| **96P/Machholz 1** | 3.03 | 0.959 | 58.1 | 0.124 | 6.8 |

---

[4]The laboratory analogue studies of Licandro *et al.* 2007 suggest an additional possible bluening mechanism beyond sublimative removal of red refractory organics. While heavy refractory organics like tholins and graphite are very red, amorphous carbon species like carbon black and soot are bluish (albeit also very dark and absorbing). Destructive pyrolization of heavy organics in vacuum, similar to how carbon fibers are industrially formed from CHON polymers in inert atmospheres on Earth, can drive off $H_2O$, $H_2S$, CO, and $N_2$, leaving behind an almost pure carbon "carbonized" soot-like residue.

Finally, since the submission of this Note, Sheppard *et al.* (2021) have announced the discovery of an asteroid on a short period orbit with an even smaller perihelion distance than Phaethon, designated as 2021 PH27. Thus two other small rocky airless bodies are now known with similarly small orbital periods and perihelion distances: asteroid (155140) 2005 UD (Devogèle *et al.* 2020, MacLennan *et al.* 2021) and asteroid 2021 PH27 (Sheppard *et al.* 2021). All of the arguments made for Phaethon in this work and in Masiero *et al.* (2021) should hold for 2021 PH27 and 2005 UD as well, so we predict that further spectrophotometric observations of these objects will show them to be bluish in color with a significant Na and Fe/Si/O coma surrounding them near perihelion. This prediction is supported by the results of our searching through the associated cometary literature to find the reports of an unusually blue and C2/C3/CN depleted nucleus for comet 96P/Machholz 1 (Eisner *et al.* 2019).

**IV.     Conclusions**. In summary, in this Note we argue that Phaethon's near-Sun perihelion, its mass-loss activity, and its rocky composition are linked to its highly bluened surface, i.e. that thermally driven sublimation at ~1100K creates "rock comet" behavior with loss rate of $Q_{gas}$ ~ $10^{22}$ mol/sec of known reddening agent Fe, and how any small airless rocky body on a similar orbit should also be active and bluened. Our sublimative Fe/Pyroxene mass loss model can be tested in Phaethon-like objects by (1) determining the time of onset of outgassing activity; (2) searching for Fe gas in a coma surrounding the object, and then closer in O and SiO, the products of destructive sublimation of pyroxene rock; and (3) determining the surface reflectance spectrum to be consistent with a mix of CAIs + FeOx + olivine + inorganic amorphous/graphitic carbon.

**V.     Acknowledgements.**  C.M. Lisse was supported by the New Horizons mission project and J.K. Steckloff was supported by NASA grants 80NSSC20K0267 and 80NSSC19K1313 for the analyses reported in this manuscript. The authors would like to thank Dr. Ron Vervack for sharing his pre-publication results concerning the surface temperature structure of Phaethon near perihelion, which help to verify our higher temperature models and those of Maseiro *et al.* 2021. They would also like to acknowledge N. Samarasinha for the suggestion of looking into the colors of periodic sungrazing comets, and D. Jewitt for inspiration and his persistent study of Phaethon over decades. Finally, the authors acknowledge the significant role that the suggestions for improvement from two anonymous reviewers made in improving the final version of this work.

## **VI.    References.**

# VII. Figures.

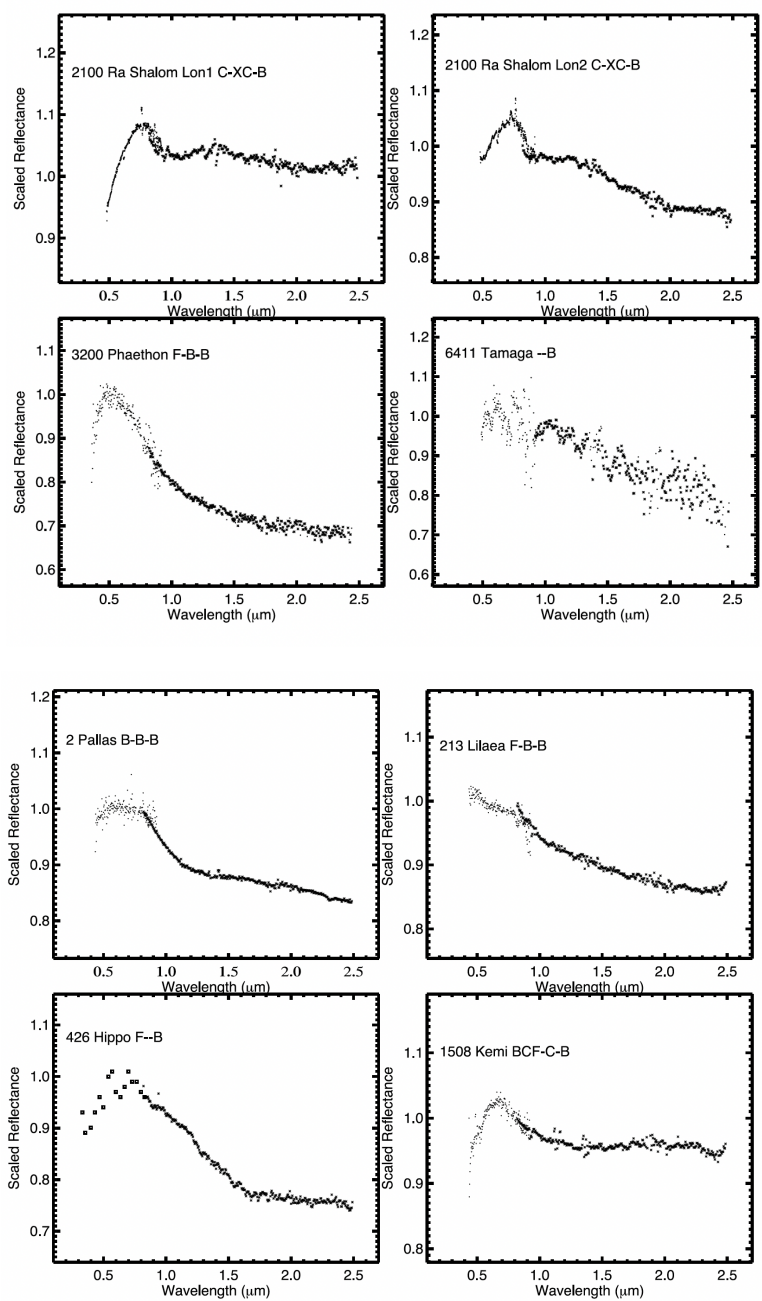
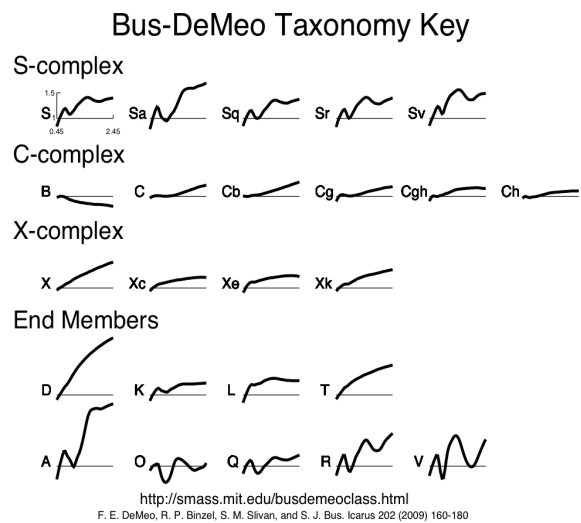

**Figure 1:** - *(Left)* Reflectance spectrum comparison for seven B-type asteroids, showing how extraordinarily **blue** Phaethon is with its 30% increase in relative reflectivity going from 2.5 down to 0.5 um, after Clark *et al.* 2013. (Typical asteroid color trends are **red** and run 5-10% amplitude over this wavelength range.) Each asteroid is shown relative to its own normalized reflectance at 0.55 um. Next to the name are the taxonomic designations of the asteroid in three different systems as follows: Tholen-Bus-Bus DeMeo. The best meteorite spectral matches found by Clark *et al.* (2013) are the highly thermally processed and oxidized CK ALH85002 and EET92002 meteorites, and the best laboratory analogue matches to the spectra are mixes of silicates and inorganic carbon (Licandro *et al.* 2007). *(Above)* Schematic representation of the Bus-DeMeo asteroid VISNIR spectral reflectance taxonomy, showing that all asteroids have reddish spectra except for the rare B-types. After DeMeo *et al.* 2009.

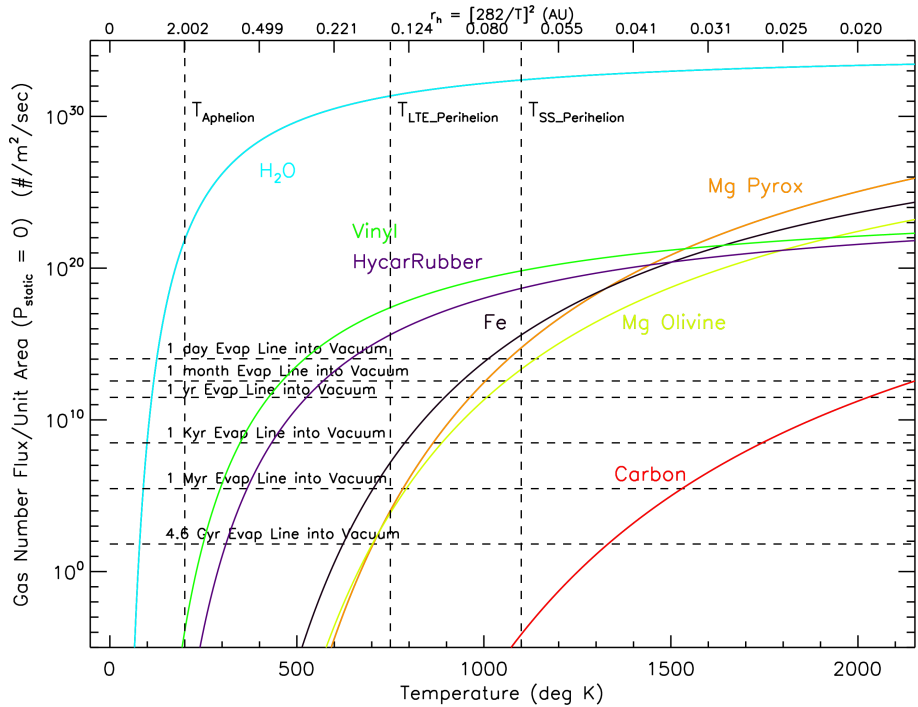

**Figure 2: Gas molecule loss rate and stability vs temperature curves of typical rocky solids (water ice, complex organics, iron, olivine/pyroxene rock, carbon) found in asteroids.** The dashed horizontal lines denote the levels of loss rate above which the solid sublimes and disappears into vacuum in 1 day, 1 month, 1 year, 1 Kyr, 1 Myr, and 4.65 Gyr (the age of our solar system), and the dashed vertical lines the LTE temperature at Phaethon's aphelion ($T_{APH}$), the LTE temperature at perihelion ($T_{LTE}$), and the sub-solar point temperature at perihelion $T_{SS}$). From the curves and dashed line limits, it is clear that **water ice is never stable as a solid throughout Phaethon's orbit. Complex organics (as represented by Vinyl and Hycar Rubber) are stable until Phaethon gets within ~0.33 AU**, the orbital distance of Mercury, or about 10% of its orbit. **Fe, by contrast, is stable until Phaethon is inside ~0.15 AU, close to its perihelion distance.** Solid rock, is always stable accept during closest approach to the Sun at local noon on the surface. Amorphous and graphitic carbon are always stable. Water ice data & format from Lisse *et al.* 2021 and references therein; complex organic Vinyl and Hycar Rubber data from Jensen 1956; MgSiO$_3$ (Mg-pyroxene) curves from Ackerman & Marley 2001 and 2013's fit of Barshay & Lewis 1976's data; Mg$_2$SiO$_4$ (Mg-olivine) data from Costa *et al.* 2017; Fe data from Alcock *et al.* 1984; C (Carbon) data from Darken & Gurry 1953.

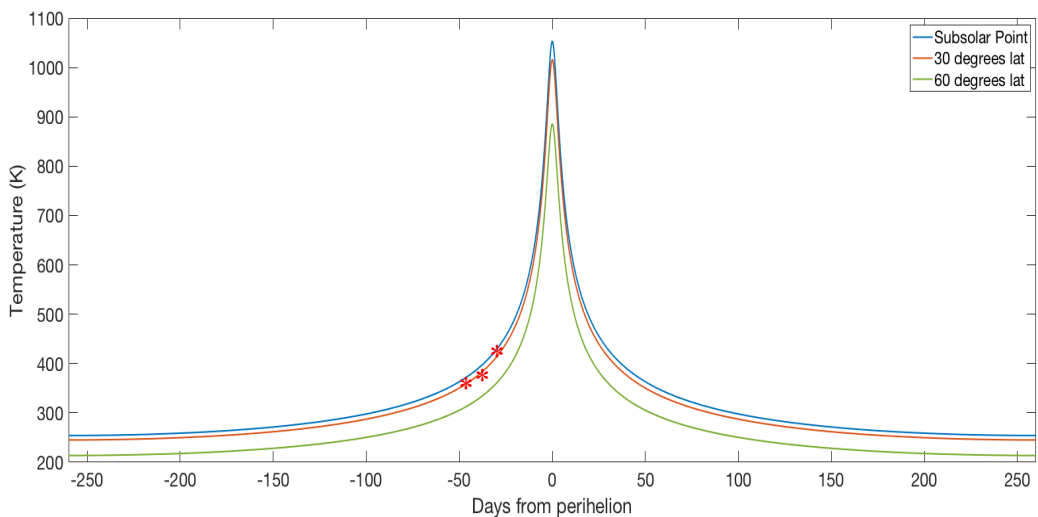

**Figure 3: Phaethon surface temperatures at the sub solar point and at 30 and 60 degrees of latitude and the sub-solar longitude for the solid species shown in Fig. 1.** The hottest points on Phaethon, near local Noon, climb above 1000 K for a few days near perihelion every orbit. Red Stars : 30 deg latitude model temperature estimates from rotationally resolved near-infrared spectroscopy of Phaethon's surface by R. Vervack *et al.* (2021).

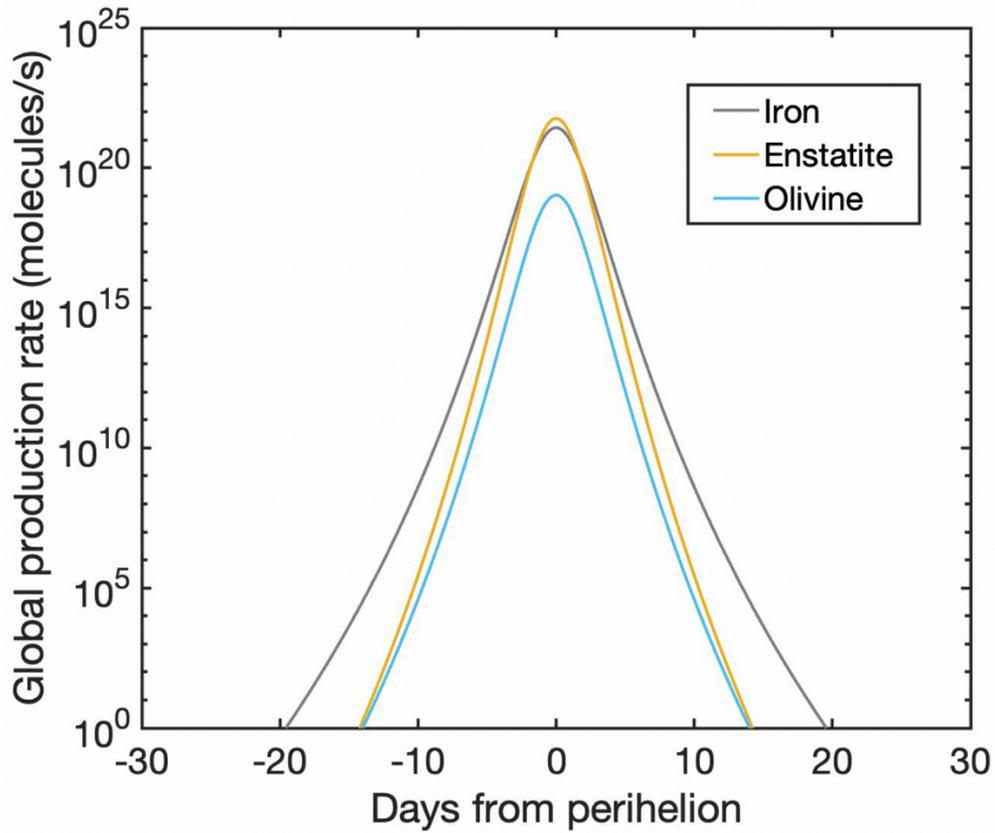

**Figure 4: Predicted total body outgassing rate** vs heliocentric distance for a 2.9 km radius spherical Phaethon using the MaMOS model (Springmann *et al.* 2019; Steckloff *et al.* 2020, 2021). Parameters assumed for Phaethon are Bond Albedo = 0.043, Emissivity ~ 0.9, beaming parameter = 0.9, 3.6 hr rotation period, thermal inertia = 600 J m$^{-2}$ K$^{-1}$ s$^{-1/2}$, orbital eccentricity = 0.88990, perihelion distance = 0.13998 AU, and orbit semimajor axis = 1.2714 AU. Iron curve is in black, pyroxene in gold, olivine in aqua. The similarity in the iron and pyroxene sublimation rates near perihelion is evident. $P_{sat}$ vs. T relations used are the same as in Fig. 1: MgSiO$_3$ (Mg-pyroxene) curves from Ackerman & Marley 2001's fit of Barshay & Lewis 1976's data; Mg$_2$SiO$_4$ (Mg-olivine) data from Costa *et al.* 2017; Fe data from Alcock *et al.* 1984.